\documentclass[twocolumn]{aastex62}
\usepackage{xcolor,soul}
\usepackage{hyperref}
\providecommand{\bjdtdb}{\ensuremath{\rm {BJD_{TDB}}}}

\providecommand{\msun}{\ensuremath{\,M_\Sun}}
\providecommand{\rsun}{\ensuremath{\,R_\Sun}}
\providecommand{\lsun}{\ensuremath{\,L_\Sun}}
\providecommand{\mj}{\ensuremath{\,{\rm M_J}}}
\providecommand{\rj}{\ensuremath{\,{\rm R_J}}}

\providecommand{\mst}{\ensuremath{\,{\rm M_\odot}}}
\providecommand{\rst}{\ensuremath{\,{\rm R_\odot}}}

\providecommand{\arcsec}{$^{\prime \prime}$}
\graphicspath{{./}{figures/}}

\begin{document}

\title{New substellar discoveries from \textit{Kepler} and \textit{K2}: Is there a brown dwarf desert?}

\author{Theron W. Carmichael}
\affil{\rm Harvard University,
Cambridge, MA 02138 \\ email: \rm \href{mailto:tcarmich@cfa.harvard.edu}{tcarmich@cfa.harvard.edu}}
\affil{\rm Center for Astrophysics $|$ Harvard \& Smithsonian, 60 Garden Street, Cambridge, MA 02138}

\author{David W. Latham}
\affil{\rm Center for Astrophysics $|$ Harvard \& Smithsonian, 60 Garden Street, Cambridge, MA 02138}

\author{Andrew M. Vanderburg}
\affil{\rm University of Texas at Austin, Austin, TX 78712}
\affil{\rm NASA Sagan Fellow}

\begin{abstract}
\noindent We present the discoveries of a brown dwarf and a low mass star from the \textit{Kepler} and \textit{K2} missions. The newly discovered brown dwarf is EPIC 212036875b and the low mass star is KOI-607b. EPIC 212036875b has a mass of $ M_{b}=52.3\pm 1.9\mj$, a radius of $ R_{b}=0.874\pm 0.017\rj$, and orbits its host star  in $ P=5.169885 \pm 0.000027$ days. Its host star is a late F-type star with $ M_\star=1.288\pm 0.065\mst$, $ R_\star=1.498\pm 0.025\rst$, and $ T_{\rm eff}=6238 \pm 60$K. KOI-607b has a mass of $ M_{b}=95.1\pm 3.4\mj$, a radius of $ R_{b}=1.089\pm 0.089\rj$, and an orbital period of $ P=5.89399148 \pm 0.00000060$ days. The primary star in the KOI-607 system is a G dwarf with $ M_\star=0.993\pm 0.052\mst$, $ R_\star=0.915\pm 0.031\rst$, and $ T_{\rm eff} = 5418\pm 87$K. We also revisit a brown dwarf, CWW 89Ab, that was previously published by \cite{nowak17} (under the designation EPIC 219388192b). CWW 89Ab is one of two known transiting brown dwarfs associated with a star cluster, which illustrates the need for more brown dwarfs with accurate masses and radii and reliable age determinations to test theoretical models. We find that the newly discovered brown dwarf, EPIC 212036875b, falls in the middle of the so-called ``brown dwarf desert", indicating that EPIC 212036875b is either a particularly rare object, or the brown dwarf desert may not be so dry after all.
\end{abstract}

\keywords{brown dwarfs – techniques: photometric – techniques: radial velocities – techniques: spectroscopic}

\section{Introduction} \label{sec:intro}
Brown dwarfs are typically defined as objects that are massive enough to sustain deuterium fusion but not massive enough to fuse hydrogen in their cores. This arbitrarily places a lower mass cutoff at 13 Jupiter masses ($\mj$) that separates planets from brown dwarfs (BDs). Although this mass cutoff is physically motivated by a distinct process (deuterium fusion), it leaves ambiguity on how these objects form. Are objects above 13$\mj$ somehow inhibited from forming like giant planets do? Placing a cutoff at 13$\mj$ can imply BDs and giant planets form differently, which may not be the case for all BDs in the mass range 13-80$\mj$. Considering this, we ought to explore whether or not there is a mass at which a change in formation mechanism occurs and use this as the cutoff between giant planets and BDs.

To take an approach more focused on formation mechanisms of BDs, we examine the BD population by measuring the masses and radii of those BDs that orbit main sequence stars. In measuring these fundamental properties of BDs as well as their orbital characteristics, we can compare them to substellar models that motivate the underlying physics of these objects and construct a story of the evolutionary histories of these BDs.

Transiting BDs are particularly special as they provide us the opportunity to characterize a BD's mass and radius well. However, fewer than a couple dozen transiting BDs have been studied \citep{ma12, nowak17, bayliss16, csizmadia16, irwin18, johnson11_bd}. A feature of the BD population as a whole is the ``brown dwarf desert". This term describes the observed lack of BD companions within 3AU to main sequence stars \citep{bdd2000}. As \cite{ma12} argue, the ``driest" part of this desert is the mass range of $35$-$55\mj$ and periods shorter than 100 days. They suggest that this gap is indicative of two distinct BD populations that result from different formation mechanisms. \cite{ma12} claim that BDs observed below $M=42.5\mj$ form in a process similar to gas giant planets while BDs more massive than this form like low-mass stars. 

At the time, this gap between $35$-$55\mj$ seemed sparsely populated enough to support the claim of two populations, even if only considering transiting BDs with well-characterized radii and masses. However, because of the small sample size \cite{ma12} had to work with, more detections of BDs are needed to confirm the existence of this depleted region. With so few short-period transiting BDs, we cannot reliably apply statistics designed for large samples of data, so we need to more thoroughly populate this region of mass-period space to verify if a ``depleted region" exists in the brown dwarf desert and what it may reveal about the origins of the BD population. This may show a more convincing trend in the BD mass distribution and start to uncover what distinguishes BDs from planets and stars besides their ability to fuse deuterium.

Any new BDs are also useful in testing the substellar evolutionary models developed by \cite{baraffe03}. In particular, the mass and radius of a BD can be directly compared to substellar isochrones to derive its age. The transit method is particularly sensitive to the BDs in the depleted region due to their short periods and relatively large transit depths for main sequence G and F type host stars. We can use these light curves and estimates of the star's properties from models to measure the companion's radius. The host star's radial velocity (RV) is measured through follow up spectra. The mass of the companion is calculated based on these RVs, parameters from the light curve, and models estimating the host star's mass. RVs also provide a sense of the eccentricity of the companion and when this orbital information is combined with the mass and radius of the companion, we have some basic clues into the history of the object.

Here we report the discovery and characterization of EPIC 212036875b. This is a BD is in the middle of the depleted region highlighted by \cite{ma12} and is one of the first known transiting BD around 50$\mj$ in such a short period. This makes EPIC 212036875b a rare ``oasis" in the brown dwarf desert. We also revisit the brown dwarf CWW 89Ab and present a new discovery of a low mass star, KOI-607b. All three of these objects are useful in benchmarking stellar and substellar evolutionary models that we discuss in later sections. Section \ref{sec:observations} gives details on the light curves and spectra that were obtained for this study. Section \ref{sec:analysis} describes the analysis techniques used to derive the host star and companion properties. Section \ref{sec:conclusion} contains discussion of the implications of these new discoveries and what we may look forward to as the TESS mission continues to release new observations.

\section{Observations}\label{sec:observations}
The light curves for KOI-607b, EPIC 212036875b (referred to hereafter as EP212b), and CWW 89Ab are from the \textit{Kepler} and \textit{K2} missions. CWW 89Ab was first published as a transiting BD in the star cluster Ruprecht 147 by \cite{nowak17}, who use the designation EPIC 219388192b. KOI-607b was roughly estimated to be a low mass star based on 2 reconnaissance spectra by the SOPHIE team \citep{santerne12}, but no orbital eccentricity was obtained.

\subsection{Contamination from nearby sources}
\cite{nowak17} took high contrast images with the Subaru/IRCS+AO188 instrument and found no nearby stellar companions that may cause noticeable contamination for CWW 89Ab. However, \cite{cww89a} report the detection of CWW 89B, which is an M-dwarf companion at a projected separation of 25AU from CWW 89A. In this work, we do not present any high contrast or adaptive optics imaging of KOI-607 or EP212. We do check for nearby sources using Gaia DR2\footnote{Source of Gaia DR2 results: \url{http://vizier.u-strasbg.fr/viz-bin/VizieR-3?-source=I/345/gaia2}}. According to Gaia DR2: EP212 shows no companions brighter than a magnitude of G=19.70 within $10\arcsec$, KOI-607 shows no companions within $10\arcsec$, and CWW 89A shows 3 stars within $10\arcsec$ that are all fainter than G=18.47, which is at least 6.11 magnitudes fainter than CWW 89A (G=12.36). Note that CWW 89A is a member of a star cluster, so finding companions within $10\arcsec$ is not too surprising.

\subsection{\textit{Kepler} and \textit{K2} light curves}\label{sec:lightcurves}
The \textit{Kepler} space telescope detected a total of 118 transits for KOI-607b, 14 transits for EP212b, and 15 transits for CWW 89Ab. KOI-607 is the faintest target of these three at V=14.6. EP212 and CWW 89A are V=11.0 and V=12.5, respectively. The light curves for EP212, CWW 89A, and KOI-607 are shown in Figures \ref{fig:ep219_rv}, \ref{fig:ep212_rv}, and \ref{fig:koi_rv}. EP212b is from \textit{K2} Campaign 16 and CWW 89Ab is from Campaign 7. 
 
These targets were initially chosen based on the companion properties that were derived from the host star light curves. Specifically, we searched for light curves that indicated: 1) a companion object's radius to be approximately 1 Jupiter radius ($\rm R_J$), 2) an orbital period on the order of 30 days or shorter, 3) a host star bright enough (V$<$15) for the 1.5m telescope that the Tillinghast Reflector Echelle Spectrograph (TRES) is installed onto to perform follow up spectroscopic observations. The light curve of KOI-607 was directly downloaded from the Mikulski Archive for Space Telescopes (MAST)\footnote{Direct link to light curve of KOI-607b: \url{https://archive.stsci.edu/pub/kepler/lightcurves/0054/005441980/}} and flattened by dividing out the median-smoothed flux. The light curves for EP212b and CWW 89Ab were taken from the MAST K2SFF archive \citep{vanderburg14}\footnote{K2 light curves from MAST: \url{https://archive.stsci.edu/prepds/k2sff/}}. After detecting the transits for these BDs, we re-derived the K2 systematics correction for these two objects by simultaneously fitting the K2 roll systematics with the transit shape and stellar variability (following \cite{vanderburg16}). We divided away the best-fit spline describing the stellar variability from our model to flatten the light curves. 

\subsection{TRES spectra}\label{sec:spectra}
The spectra for KOI-607, EP212, and CWW 89A were taken with the TRES instrument on Mt. Hopkins, Arizona. The spectrograph has a resolution of R=44,000 and covers wavelengths from 390nm to 910nm. CWW 89A has 18 TRES spectra that were taken in 2015 and 2016 with exposure times ranging from 1200s to 1700s and S/N (signal-to-noise per resolution element) ranging from 22 to 34 (except for one point near phase 0.86, which has an exposure time of 400s and S/N of 6.8; this spectrum is not used to derive stellar parameters). EP212 has 14 TRES spectra taken in 2018 with exposure times ranging from 400s to 1800s and S/N ranging from 22 to 45. KOI-607 also has 14 TRES spectra and these were taken in 2014 (except for one point at phase 0.80 that was taken in 2018) with exposure times ranging from 1800s to 3600s and S/N ranging from 13 to 17.

We use multiple orders in each echelle spectrum to measure the RV at each phase. We visually omit individual orders with poor S/N and manually remove obvious cosmic rays. This leaves us with a wavelength range of 458nm-606nm, or 18 echelle orders, for our objects. Each order is cross-correlated with the corresponding order of a spectrum of the target star for relative velocity measurements. This yields a velocity shift and the average of this shift over all 18 orders is taken as the RV at each respective phase of the orbit. We choose a stellar template spectrum that is of the target star for these relative RVs. The chosen target star template spectrum is the one with the highest S/N.

\section{Analysis} \label{sec:analysis}
\subsection{Stellar Parameter Classification}
We use the stellar parameter classification (SPC) software package by \cite{spc} to derive effective temperature ($T_{\rm eff}$), metallicity, surface gravity ($\log{g}$), and the projected stellar equatorial velocity ($v\sin{i}$) from the spectra of our objects. We use SPC on a co-added spectrum for each object. We do not co-add any spectrum with S/N$<$15. When using SPC, we use the 5030\AA-5320\AA wavelength range (centered on the Mg b triplet) for each TRES spectrum. SPC is not used to measured RVs.

\subsection{Modeling with EXOFASTv2}\label{sec:exofast}
\begin{figure}[!ht]
\centering
\includegraphics[width=0.40\textwidth, trim={0.0cm 0.0cm 1.0cm 0.0cm}]{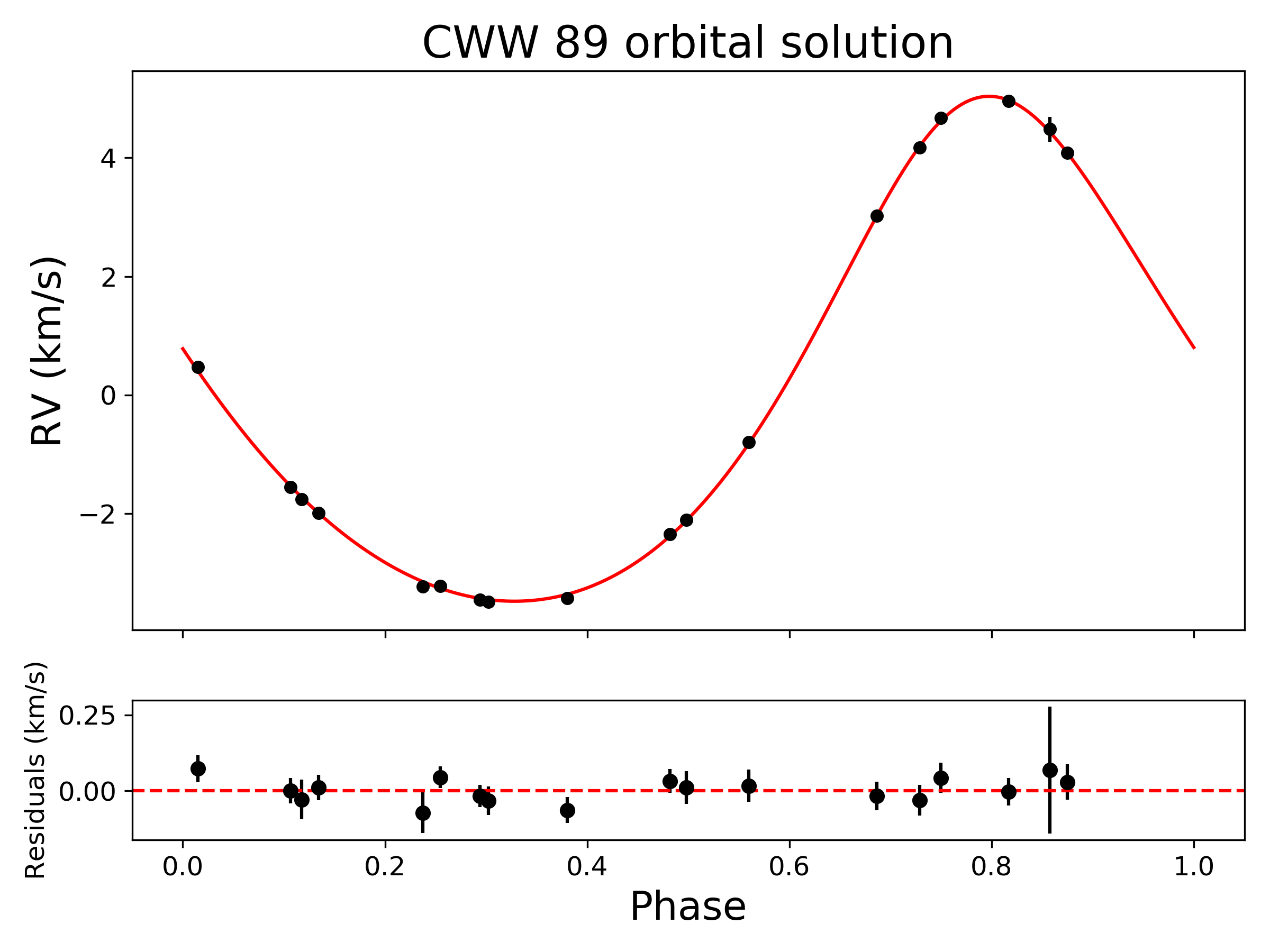}
\includegraphics[width=0.40\textwidth, trim={0.0cm 0.0cm 2.5cm 0.0cm}]{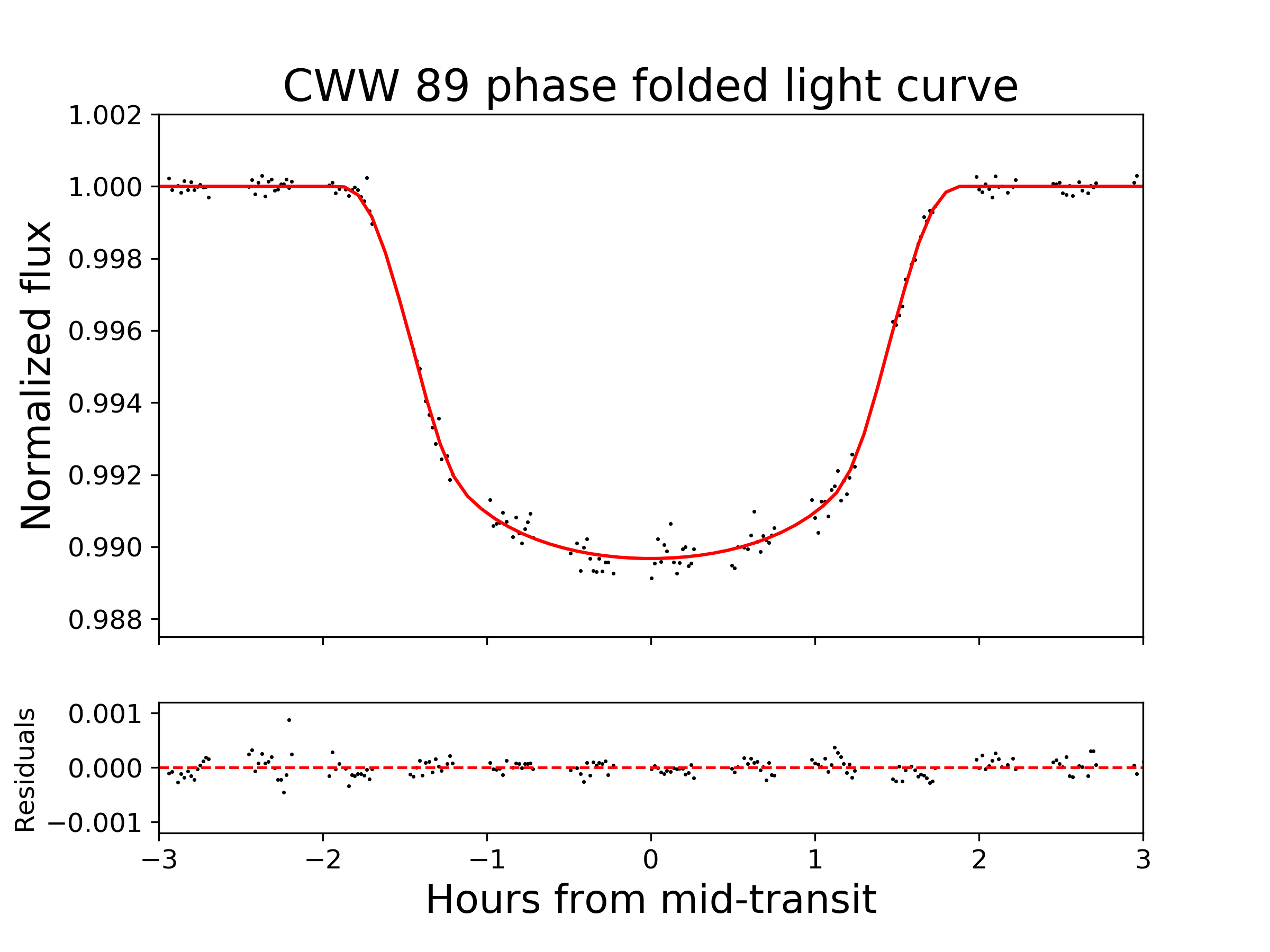}
\caption{Top panel: TRES multi-order relative radial velocities of CWW 89A. The EXOFASTv2 orbital solution is plotted in red. Bottom panel: \textit{Kepler} light curve with EXOFASTv2 transit model in red.}\label{fig:ep219_rv}
\end{figure}

\begin{figure}[!ht]
\centering
\includegraphics[width=0.40\textwidth, trim={0.0cm 0.0cm 1.0cm 0.0cm}]{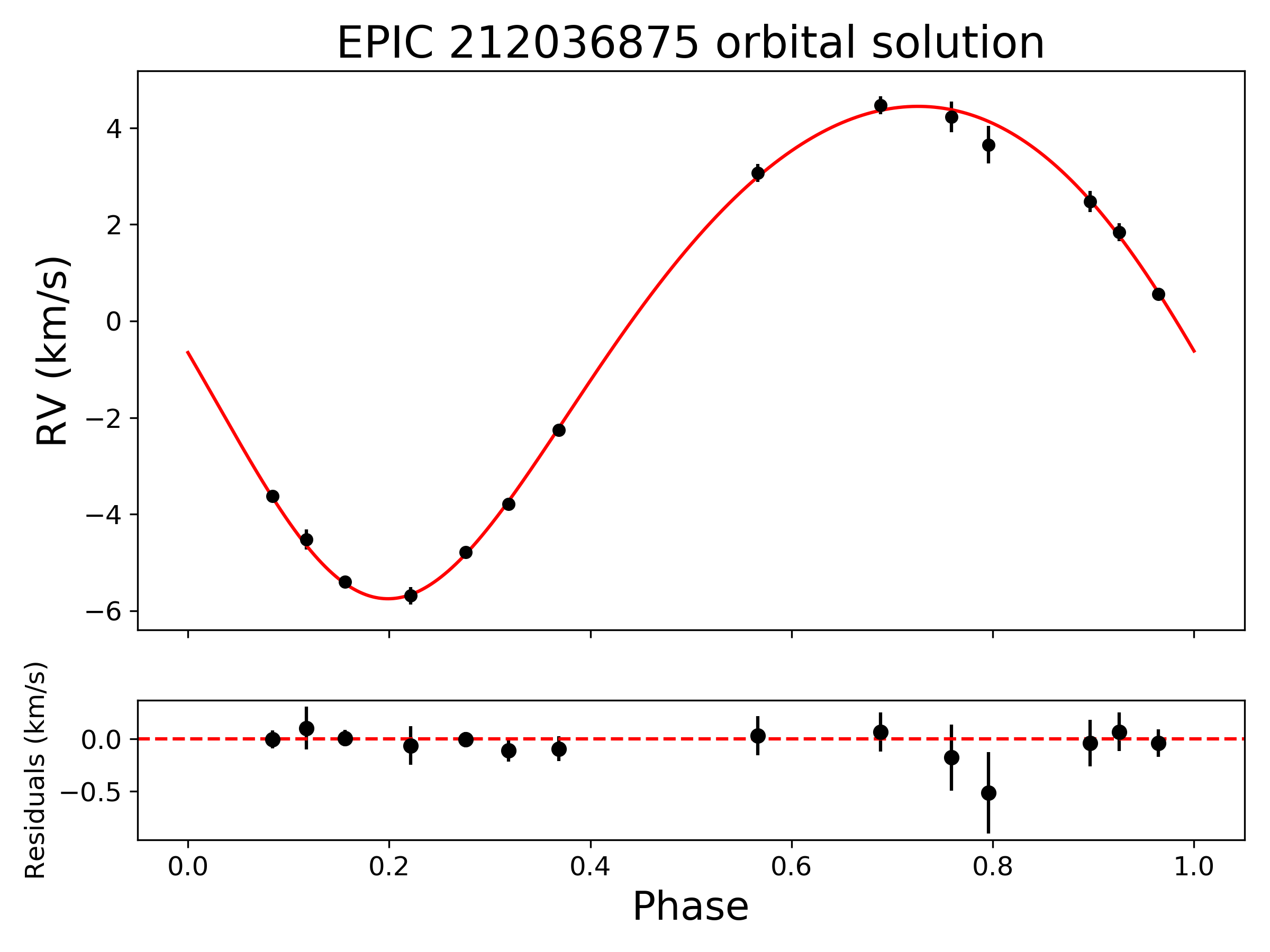}
\includegraphics[width=0.40\textwidth, trim={0.0cm 0.0cm 2.5cm 0.0cm}]{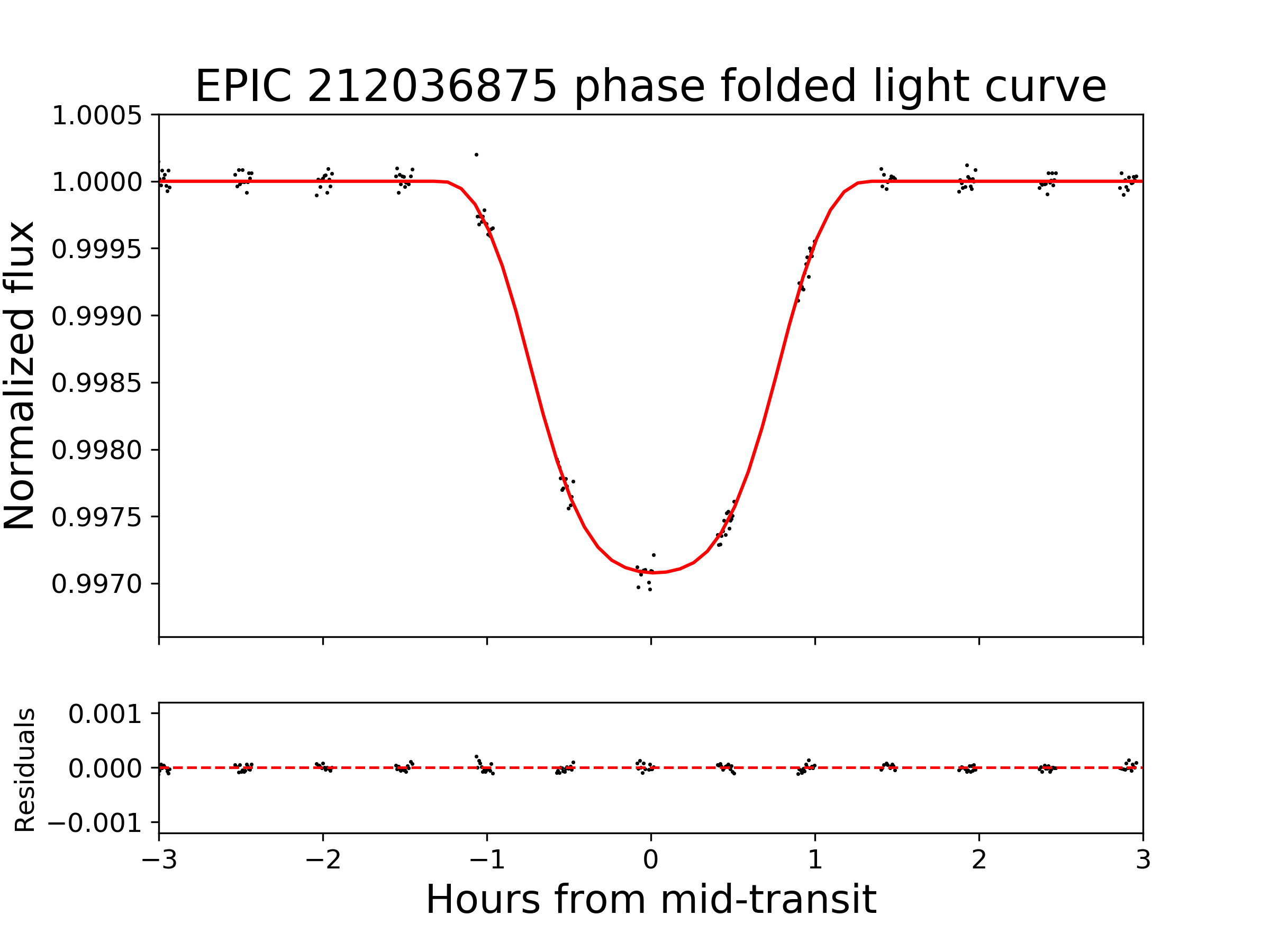}
\caption{Top panel: TRES multi-order relative radial velocities of EPIC 212036875 with EXOFASTv2 orbital solution plotted in red. Bottom panel: \textit{Kepler} light curve with EXOFASTv2 transit model in red.}\label{fig:ep212_rv}
\end{figure}

\begin{figure}[!ht]
\centering
\includegraphics[width=0.40\textwidth, trim={0.0cm 0.0cm 0.0cm 0.0cm}]{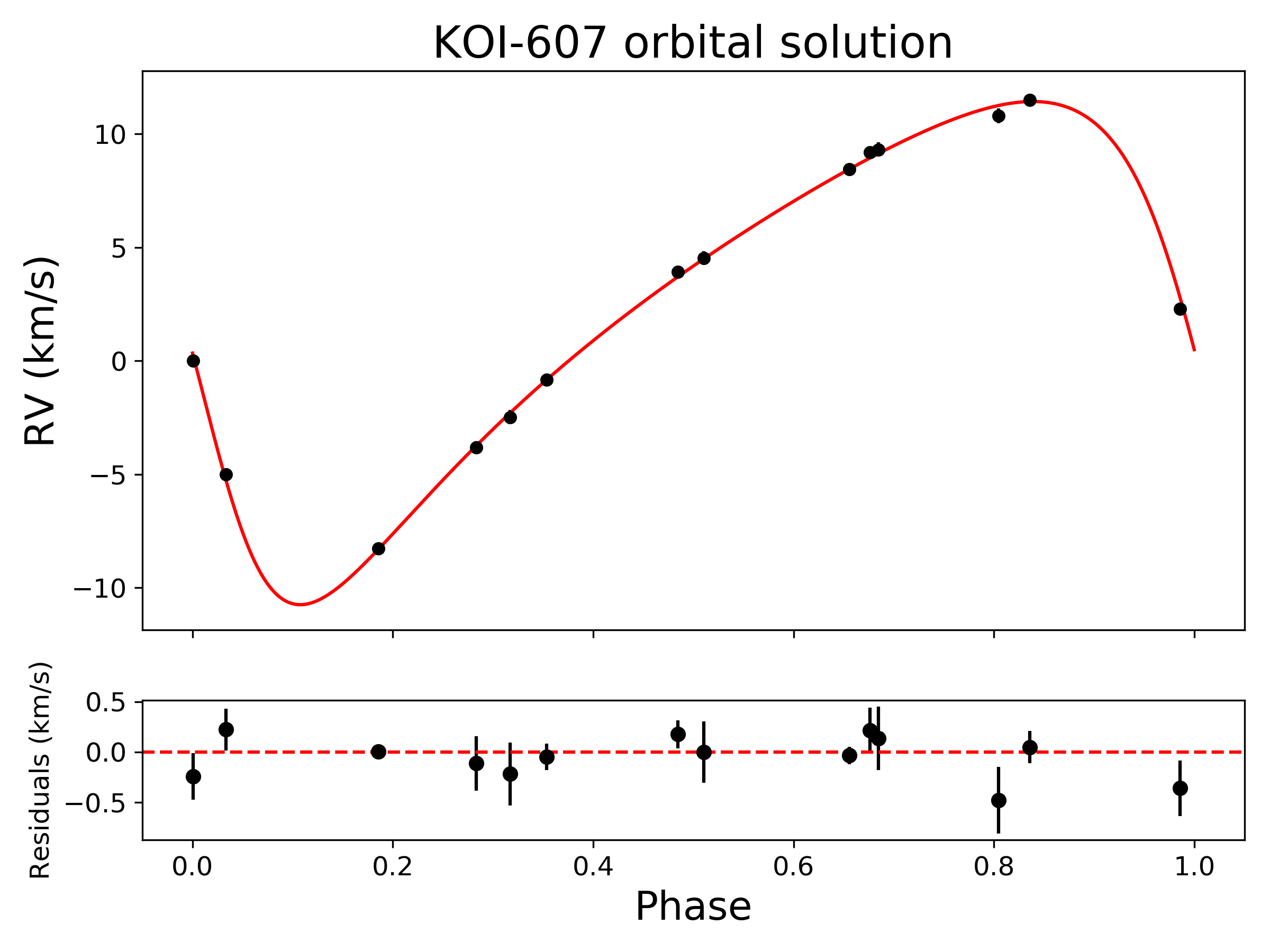}
\includegraphics[width=0.40\textwidth, trim={0.0cm 0.0cm 1.5cm 0.0cm}]{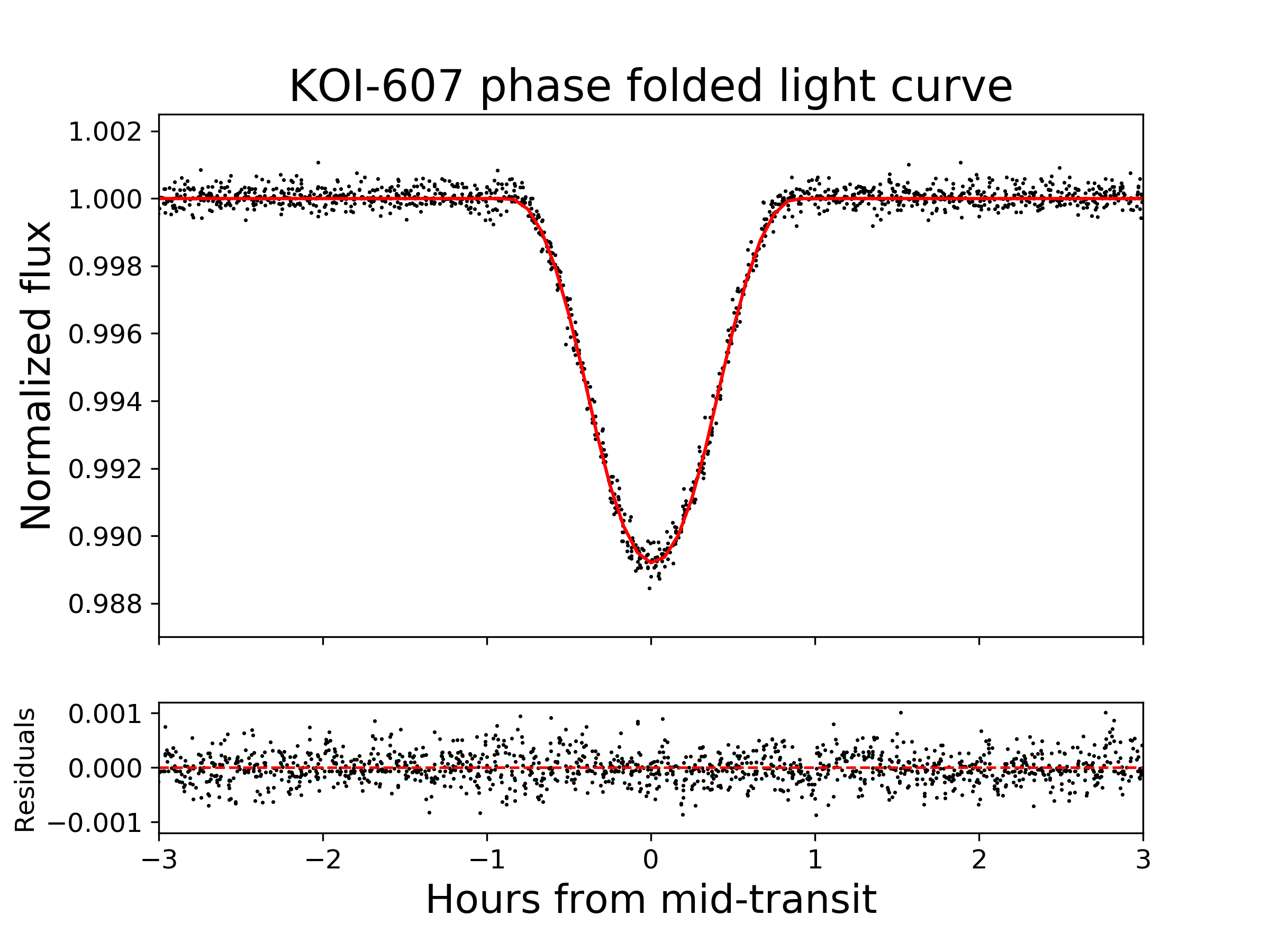}
\caption{Top panel: TRES multi-order relative radial velocities of KOI-607 with EXOFASTv2 orbital solution plotted in red. Bottom panel: \textit{Kepler} light curve with EXOFASTv2 transit model in red.}\label{fig:koi_rv}
\end{figure}

The masses and radii of the companions are derived using EXOFASTv2. A full description of EXOFAST is given in \cite{eastman13}. EXOFASTv2 uses the Monte Carlo-Markov Chain (MCMC) method. For each MCMC fit, we use N=36 (N = 2$\times n_{\rm parameters}$) walkers, or chains, and run for 50,000 steps, or links. The host star masses and radii are modelled using the MIST isochrones \citep{mist1, mist2, mist3}, which are integrated into the framework of EXOFASTv2. The resulting RV and transit fits are shown in Figures \ref{fig:ep219_rv}, \ref{fig:ep212_rv}, and \ref{fig:koi_rv}. We account for interstellar extinction, $A_V$, using the Galactic dust and reddening extinction tool from IRAS and COBE/DIRBE \footnote{Galactic dust and reddening extinction tool: \url{https://irsa.ipac.caltech.edu/applications/DUST/}} and take this $A_V$ value as an upper limit for our priors in EXOFASTv2. We also use the parallax of each host star as measured by Gaia DR2 and the SPC results for $T_{\rm eff}$ and $\rm [Fe/H]$ as starting points for our priors. The full list of free parameters we specify for each object is: period $P$, time of conjunction ($T_C$ in BJD), host star effective temperature $T_{\rm eff}$, host star metallicity $\rm [Fe/H]$, RV semi-amplitude $K$, RV relative offset $\gamma_{rel}$, interstellar extinction $A_V$, parallax, orbital inclination $i$, eccentricity $e$, and $R_B/R_\star$. The derived $T_{\rm eff}$ from EXOFASTv2 agrees well with the spectroscopic $T_{\rm eff}$ from SPC. We impose Gaussian priors on these free parameters in EXOFASTv2. The median value with 1-$\sigma$ uncertainties of the MCMC chains for each parameter is reported in Tables \ref{tab:EP219388192_MIST}, \ref{tab:EP212036875_MIST}, and \ref{tab:K00607_MIST}.

\subsection{Host star \& companion properties}\label{sec:hoststars}

\subsubsection{CWW 89A}\label{sec:ep219}
For CWW 89Ab, we derive BD and stellar parameters that agree with those reported by \cite{nowak17}. We compare these values in Table \ref{tab:compare}. A detailed list of the BD and host star properties we derive is given in Table \ref{tab:EP219388192_MIST}. CWW 89Ab is associated with the star cluster Ruprecht 147, which has an age of $2.48 \pm 0.25$ Gyr \citep{r147_age}. \cite{nowak17} find an age of $3.9 \pm 1.9$ Gyr and report a distance of $300 \pm 24$pc to the cluster, which is consistent with the Gaia DR2 distance of $ 302.7 \pm 4.2$pc.

\begin{deluxetable}{ccc}
\tabletypesize{\footnotesize}
\tablewidth{0pt}

 \tablecaption{Comparison of parameters for CWW 89Ab to \cite{nowak17} (N17). \label{tab:compare}}

 \tablehead{
 \colhead{Parameter} & \colhead{N17} & \colhead{This work}}

\startdata 
$M_b$ ($\rm M_J$) &  $36.50 \pm 0.09$ & $ 39.2 \pm 1.1$\\
$R_b$ ($\rm R_J$)&  $0.937 \pm 0.042$ & $0.941 \pm 0.019$\\
Period (days) & $5.2926$ & $5.2926$\\
$e$ & $0.1929 \pm 0.0019$ & $ 0.1891 \pm 0.0022$\\
Orbital inclination $i$ & $90.0 \pm 0.7$ & $ 88.53 \pm 0.27$\\
$M_\star$ ($\rm \mst$) & $0.99 \pm 0.05$ &  $ 1.101 \pm 0.045$\\
$R_\star$ ($\rm \rst$)&  $1.01 \pm 0.04$ &  $ 1.029 \pm 0.016$\\
$\log{g}$ &  $4.38 \pm 0.12$ &  $ 4.455 \pm 0.020$\\
$T_{\rm eff}$ (K) &  $5850 \pm 85$ &  $ 5755 \pm 49$\\
$[{\rm Fe/H}]$ & $0.03 \pm 0.08$ &  $ 0.203 \pm 0.091$\\
$v_{\rm rot} \sin{i_\star} $ (km/s) &  $4.1 \pm 0.4$ &  $5.6 \pm 0.5$\\
$P_{\rm rot}$ (days) & $12.6 \pm 2.1$ & $8.48 \pm 0.76$\\
\enddata
\vspace{-3.5cm}
\end{deluxetable}

\begin{deluxetable}{ccc}
\tabletypesize{\footnotesize}
\tablewidth{0pt}

 \tablecaption{Multi-order relative radial velocities of CWW 89A from TRES. \label{tab:ep219_rvs}}

 \tablehead{
 \colhead{$\rm BJD$ (2 450 000+)} & \colhead{RV (m/s)} & \colhead{$\sigma_{\rm RV}$ (m/s)}}

\startdata 
7289.710285 &    0.00 &   46.46\\
7294.660246 &  262.33 &   67.55\\
7304.700379 & 1498.87 &   41.97\\
7508.965702 & 7665.12 &   49.61\\
7510.965409 & 1939.65 &   41.87\\
7511.956992 &   34.72 &   36.52\\
7512.950940 & 1143.93 &   39.48\\
7514.939177 & 7970.80 &  209.07\\
7523.949521 & 2698.23 &   53.00\\
7524.955012 & 8161.81 &   49.46\\
7526.899520 & 1729.28 &   65.11\\
7528.912716 & 1380.25 &   53.94\\
7529.911743 & 6509.29 &   46.46\\
7530.909333 & 7572.51 &   58.17\\
7532.918567 &  265.27 &   35.97\\
7535.892641 & 8444.05 &   45.24\\
7536.942386 & 3957.09 &   43.99\\
7538.875832 &   60.76 &   42.96\\
\enddata
\vspace{-3.5cm}
\end{deluxetable}

\subsubsection{EPIC 212036875}\label{sec:ep212}
EP212b is a newly discovered BD with a mass of $ M_{b}=52.3 \pm 1.9 \mj$, a radius of $ R_{b}= 0.874 \pm 0.017 \rj$, an orbital period of $ P=5.169885 \pm 0.000027$ days, and an eccentricity $ e=0.1323 \pm 0.0042$. The host star properties are $ M_\star = 1.288 \pm 0.065 \mst$, $ R_\star=1.498 \pm 0.026 \rst$, $ T_{\rm eff}=6238 \pm 60$K, $ \log{g}=4.196 \pm 0.020$, and [Fe/H] $ = 0.007 \pm 0.096$. A full list of the BD and host star properties is given in Table \ref{tab:EP212036875_MIST}.

\begin{deluxetable}{ccc}
\tabletypesize{\footnotesize}
\tablewidth{0pt}

 \tablecaption{Multi-order relative radial velocity measurements of EPIC 212036875 from TRES. \label{tab:ep212_rvs}}

 \tablehead{
 \colhead{$\rm BJD$ (2 450 000+)} & \colhead{RV (m/s)} & \colhead{$\sigma_{\rm RV}$ (m/s)}}

\startdata 
8410.016273 &    0.00 &   184.38\\
8412.985125 &  9340.43 &   389.48\\
8415.947492 &  3433.55 &   118.49\\
8416.969995 &  8755.55 &   185.48\\
8417.965811 &  9922.14 &   316.15\\
8419.028375 &  6246.57 &   132.86\\
8420.018728 &   294.11 &   79.06\\
8423.997406 &  7529.63 &   184.38\\
8424.991730 &  1169.48 &   206.02\\
8426.030115 &  1904.94 &   107.71\\
8427.941099 & 10160.29 &   184.59\\
8429.017398 &  8165.64 &   223.73\\
8429.985319 &  2067.84 &   86.43\\
8430.979145 &   905.48 &   73.78\\
\enddata
\vspace{-3.5cm}
\end{deluxetable}

\subsubsection{KOI-607}\label{sec:koi607}
KOI-607b is a very low mass star with a mass of $ M_{b}=95.1 \pm 3.4 \mj$, a radius of $ R_{b}=1.089 \pm 0.089 \rj$, an orbital period of $ P=5.89399148 \pm 0.0000006$ days, and an eccentricity of $ e=0.3950 \pm 0.0091$. The host star properties that we derived are $ M_\star=0.993 \pm 0.052 \mst$, $ R_\star=0.915 \pm 0.031 \rst$, $ T_{\rm eff}=5418 \pm 87$K, $ \log{g}=4.511 \pm 0.023$, and [Fe/H] $ = 0.376 \pm 0.088$. A full list of the BD and host star properties is given in Table \ref{tab:K00607_MIST}.

\begin{deluxetable}{ccc}
\tabletypesize{\footnotesize}
\tablewidth{0pt}

 \tablecaption{Multi-order relative radial velocity measurements of KOI-607 from TRES. \label{tab:koi607_rvs}}

 \tablehead{
 \colhead{$\rm BJD$ (2 450 000+)} & \colhead{RV (m/s)} & \colhead{$\sigma_{\rm RV}$ (m/s)}}

\startdata 
6799.929541 & -3810.40 &   269.05\\
6815.858574 &  2281.54 &   276.12\\
6817.809322 & -2480.08 &   311.32\\
6818.796316 &  3911.57 &   138.46\\
6819.805144 &  8446.57 &   85.84\\
6820.866283 & 11493.66 &   160.12\\
6821.839033 &    0.00 &   228.96\\
6825.821742 &  9187.90 &   228.96\\
6827.925179 & -5012.46 &   208.13\\
6828.823062 & -8260.62 &   60.75\\
6829.813412 &  -843.18 &   131.95\\
6830.738371 &  4532.56 &   303.96\\
6831.764589 &  9304.38 &   314.15\\
8376.696330 & 10796.31 &   330.11\\
\enddata
\vspace{-3.5cm}
\end{deluxetable}

\subsubsection{COND03 evolutionary brown dwarf models}\label{subsec:cond03}
\begin{figure}[!ht]
\centering
\includegraphics[width=0.485\textwidth, trim= {0.0cm 1.0cm 1.0cm 0.8cm}]{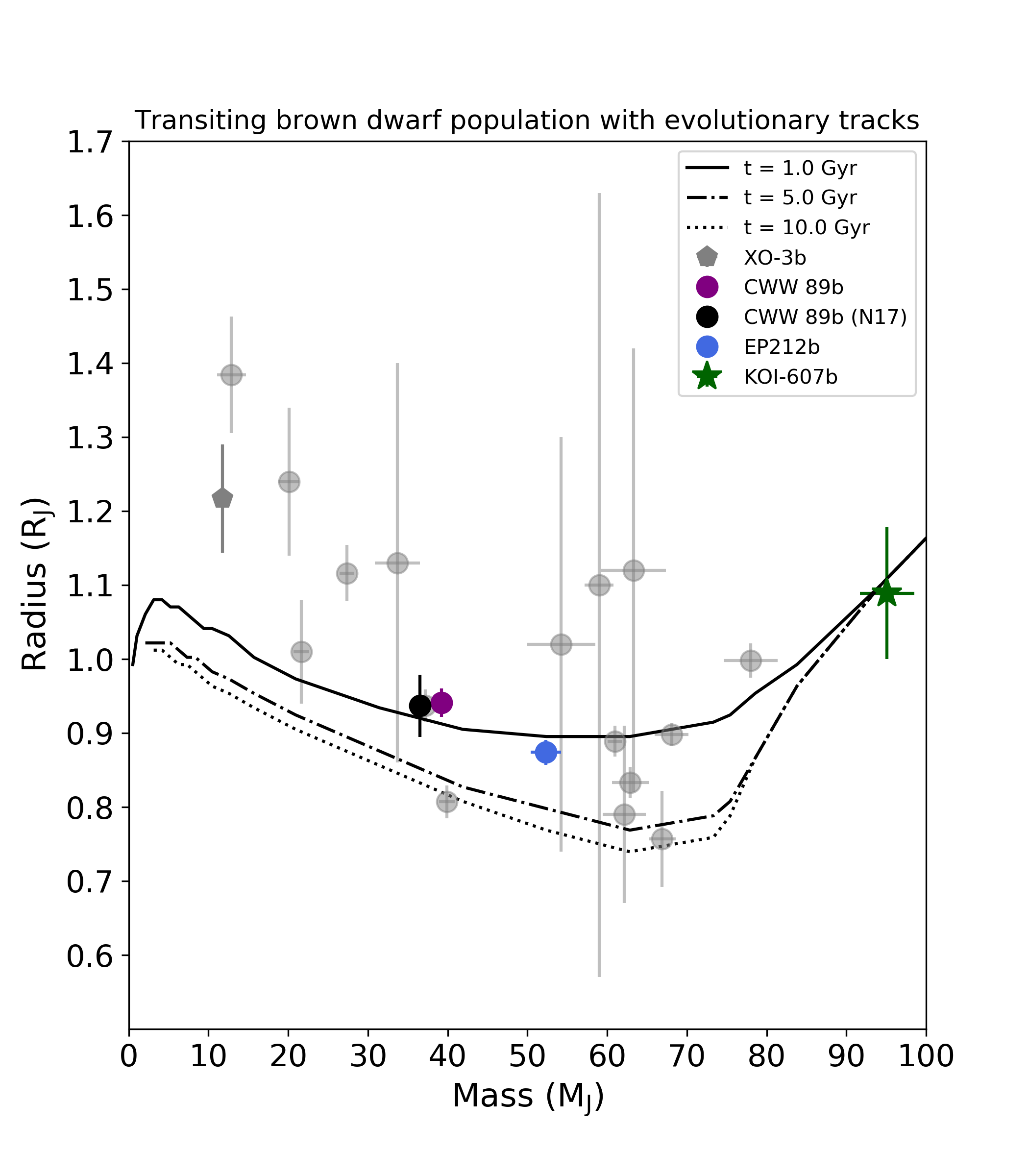}
\caption{Evolutionary brown dwarf models of mass versus radius \citep{baraffe03, saumon08} with known transiting BDs over plotted. For EPIC 212036875b, CWW 89Ab, and KOI-607b, the error bars shown are 1-$\sigma$. (COND03 models: \url{http://perso.ens-lyon.fr/isabelle.baraffe/COND03_models}).} \label{fig:mr_age}
\end{figure}

Using the masses and radii that were jointly derived with EXOFASTv2, the \textit{Kepler}/\textit{K2} light curves, and TRES RVs, we may now examine how these values compare to the evolutionary models (the COND03 models) developed by \cite{baraffe03}. In the case of CWW 89Ab, we have an independent cluster age associated with this BD ($2.50$ Gyr) and, as shown in Figure \ref{fig:mr_age}, the BD radius derived by this work and \cite{nowak17} fall on evolutionary tracks ($1.00$ Gyr) that do not match the cluster age within 1-$\sigma$ of the radius uncertainties. CWW 89Ab is one of the only BD system with a measured age (in addition to RIK 72b and AD 3116b) to test the COND03 substellar evolutionary tracks. Based only on the match to the mass-radius models, this work finds CWW 89Ab to be roughly 1 Gyr old.

To be thorough in our comparison to \cite{nowak17}, we note here that we use the same \textit{K2} observations as they do, but our reduced light curves originate from different pipelines (\cite{nowak17} use a light curve from \cite{k2_dai16}) and our spectra come from a different spectrograph. The difference in spectral analysis pipelines used may be why our $T_{\rm eff}$ value differs from that of \cite{nowak17} (see Table \ref{tab:compare}). We find an effective temperature about 100K smaller than \cite{nowak17}. We make sure to omit any low S/N spectra from the calculation of the stellar parameters, including $T_{\rm eff}$. Our effective temperature is in better agreement with that reported by \cite{cww89a} ($T_{\rm eff}=$5715K). Importantly, the radius we derive matches that derived by \cite{nowak17}, which means that we draw a similar conclusion to the age of CWW 89Ab as predicted by the \cite{baraffe03} evolutionary models. The differences in $T_{\rm eff}$ and $M_b$ between the two works do not change the conclusions about the age of the system.

With regard to the transiting BD population as a whole, it appears that the BDs less massive than $25\mj$ (except for CoRoT-3b) are over-inflated compared to the the COND03 models, though the uncertainties in radius on these objects is larger than some of the more massive BDs (Figure \ref{fig:mr_age}). The BD cooling models from \cite{baraffe03} indicate that an object with a mass on the order of $ 10\mj$ and an age greater than 500 Myr may maintain a radius of at least 1.0-1.2$\rj$. \cite{burrows07} present a discussion on radius anomalies for giant planets and show that increased internal atmospheric opacities (as a result of enhanced metallicities) of these objects may allow these objects to retain heat more efficiently and as a result, maintain larger radii. Another factor in giant planet over-inflation is a high equilibrium temperature due to a close proximity to the host star, but despite the clear separation of over-inflated hot Jupiters from their cooler counter-parts in the giant planet population \citep{thorngren17, hats70b}, it is not clear whether or not this mechanism also plays a role in the over-inflated BDs we see in Figure \ref{fig:mr_age}. What we have here are a few individual BD systems that show the radius anomalies that are present in part of the giant planet population. Indeed, it may be possible that the mechanisms driving the radius anomalies in giant planets are also responsible for the radius anomalies in these few low-mass BDs, but we do not yet have an understanding of whether or not this is truly the case.

Lastly, we also note 4 objects with some of the largest error bars for radius as AD 3116b, NLTT 41135b, CoRoT-15b, and CoRoT-33b. Of these, CoRoT-33b and NLTT 41135b are grazing transits, which explains the relatively large uncertainty in radius due to the low transit depth-to-noise ratio and degeneracy between the impact parameter and transit depth for grazing transits \citep{corot33b, irwin10}.

\section{Discussion}\label{sec:conclusion}

\subsection{The current transiting BD population}
\begin{figure}[!ht]
\centering
\includegraphics[width=0.425\textwidth, trim= {2.0cm 0.0cm 1.5cm 0.0cm}]{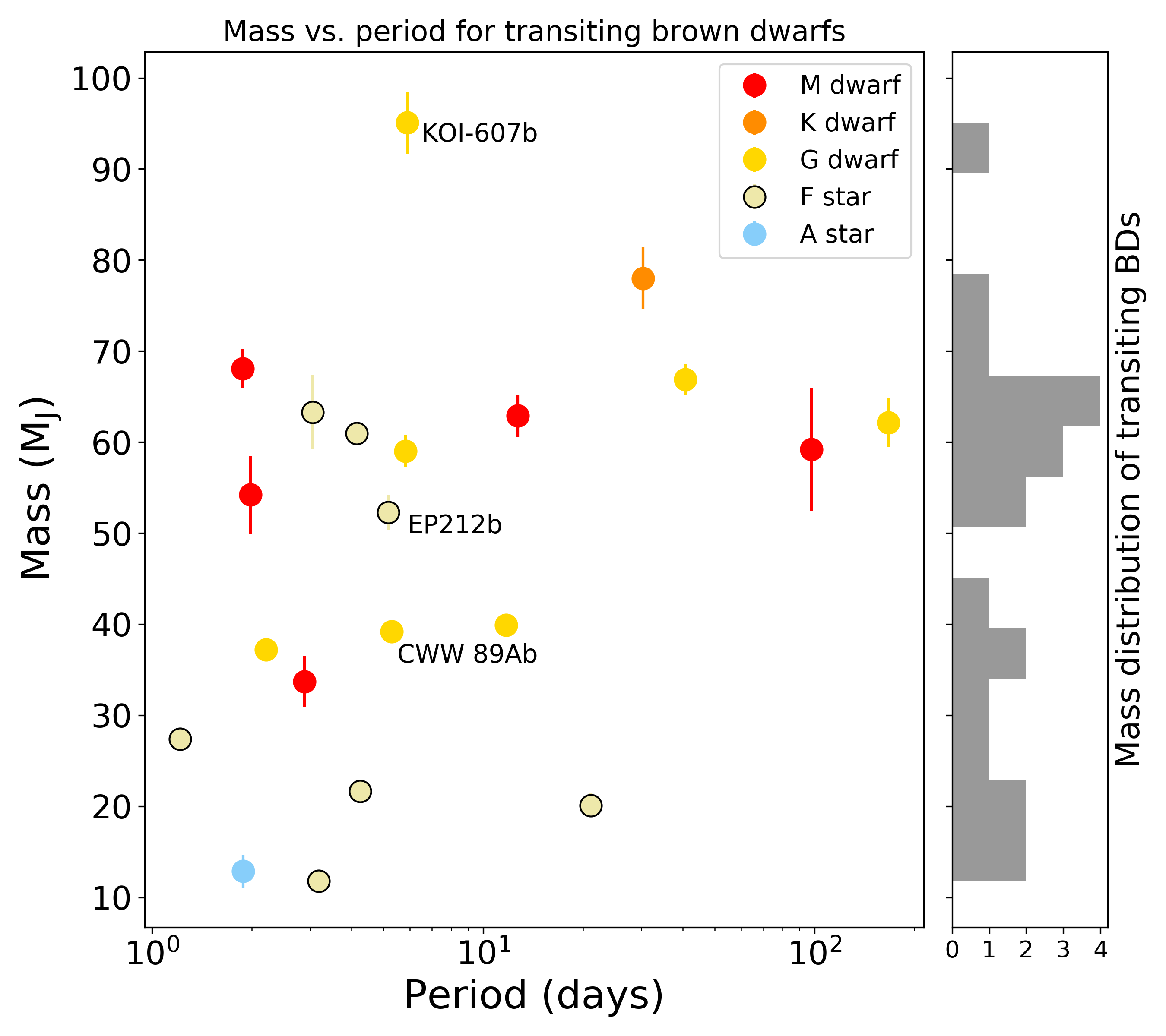}
\caption{Mass vs. Period plot showing only transiting BDs with KOI-607b for context. Host star spectral types, masses, and periods are from \cite{ma12}. The distribution of the companion masses is plotted in the histogram on the right and the colors of the points represent the spectral type of the host star in each system. There is an insufficient number of transiting BDs to determine a trend in the mass distribution or in the types of host stars for a given mass or orbital period. \label{fig:gap_hist}}
\end{figure}

Including the new BD in this work, the total number of known BDs that transit a star is 19. There is an additional binary BD system, 2M0535-05 \citep{2M0535}, bringing the total number of BDs with measured masses and radii to 21. We do not show this BD binary system in our figures. Those BDs that orbit main sequence stars are shown in Figure \ref{fig:gap_hist}. Note that the BD called RIK 72b, orbits a \textit{pre}-main sequence star \citep{david19_bd}. The list of the known transiting BDs is shown in Tables \ref{tab:bdlist} and \ref{tab:bdlist_extra}. One of these BDs, KOI-189b, has a mass of $M_b=78\mj$, placing it at the upper edge of the mass limit for BDs. Another object featured in Table \ref{tab:bdlist} is HATS-70b, which is currently the only known BD that transits an A type star \citep{hats70b}. HATS-70b has a mass of $M_b=12.9\mj$, placing it at the opposite end of the nominal BD mass range to KOI-189b. Near HATS-70b in terms of mass is XO-3b ($M_b=11.8\mj$), which is officially classified as an exoplanet. EP212b and CWW 89Ab are within the ``depleted region" of 35-55$\mj$. 

We note here that the conclusion \cite{ma12} make claiming that two populations of BDs exist (one above and one below 42.5$\mj$ in the depleted region) was based on a two-dimensional Kolmogorov-Smirnov (KS) test in period-eccentricity space for the BD population. As an astrostatistics study done by \cite{kstest} discusses, the KS test is not applicable in two or more dimensions\footnote{The two-\textit{sample} KS test determines the probability that two data sets come from the same parent distribution, however a two-dimensional KS test cannot be used ``because there is no unique way to order the points so that distances between well-defined empirical distribution functions can be computed" (from \url{https://asaip.psu.edu/Articles/beware-the-kolmogorov-smirnov-test})}.

\subsection{Circularization timescales and orbital synchronization}
We also examine the predictions for the circularization timescales of our targets. The equation for the circularization timescale $\tau_{\rm circ}$ presented in \cite{adams06} is:

\begin{equation}\label{eq:tau}
    \tau_{\rm circ} = \frac{4Q_pM_Ba^5}{63 M_\star R_B^5} \left(\frac{a^3}{GM_\star}\right)^{\frac{1}{2}}\frac{(1-e^2)^{\frac{13}{2}}}{F(e^2)}
\end{equation}

\noindent where $Q_p \approx 10^5-10^6$ is the tidal quality factor, $M_\star$ is the host star's mass, $M_B$ is the companion object's mass, $R_B$ is the companion object's radius, $a$ is the semi-major axis, and $e$ is the orbital eccentricity. Here, $F(e^2) \approx 1 + 6e^2 + \mathcal{O}(e^4)$. Since KOI-607b is a star, the tidal quality factor $Q_p$ is better approximated at a value closer to $10^6$ instead of $10^5$ \citep{qp}. In general, $Q_p$ is difficult to measure and usually only accurate to an order of magnitude \citep{qp}. In the case of CWW 89Ab, we have a more precise value of $Q_p=10^{4.5}$ from work by \cite{cww89a}, who present a detailed study of CWW 89Ab's tidal properties. Inserting the appropriate values for KOI-607b, EP212b, and CWW 89Ab, we find that the circularization timescales for all of these objects well exceed the age of each object (see Table \ref{tab:timescale}). These results are consistent with what the eccentricities of our targets imply: that each of our targets are not old enough to have surpassed their respective circularization cutoff timescale. Increasing $Q_p$ to $10^6$ for EP212b and CWW 89Ab only increases $\tau_{\rm circ}$, yielding the same conclusions drawn here.

\begin{deluxetable}{cccc}
\tabletypesize{\footnotesize}
\tablewidth{0pt}

 \tablecaption{Comparison of COND03 ages to circularization timescale $\tau_{\rm circ}$. \label{tab:timescale}}

 \tablehead{
 \colhead{Name} & \colhead{COND03 age} & \colhead{$\tau_{\rm circ}$} & \colhead{$Q_p$}}

\startdata 
EPIC 212036875b & 5 Gyr &  47 Gyr & $10^5$\\
CWW 89Ab &  1 Gyr &   7.5 Gyr & $10^{4.5}$\\
KOI-607b & 10 Gyr &  98 Gyr & $10^6$\\
\enddata
\vspace{-2.5cm}
\end{deluxetable}

To examine how well synchronized the stellar rotation and orbital periods are, we compare the quantities in Table \ref{tab:vsini}. Using SPC and stellar radii derived from EXOFASTv2, we estimate the rotation rate of each star from the Doppler line broadening in its spectra. We find that the orbits are not synchronized with the projected rotation of the respective host stars.

\begin{deluxetable}{cccc}
\tabletypesize{\footnotesize}
\tablewidth{0pt}

 \tablecaption{Comparison of stellar rotation period to orbital period. \label{tab:vsini}}

 \tablehead{
 \colhead{Name} & \colhead{$v\sin{i}$} & \colhead{$P_{\rm rot}$} & \colhead{$P_{\rm orb}$}\\  \colhead{} & \colhead{(km/s)} & \colhead{(days)} & \colhead{(days)}}

\startdata 
EPIC 212036875 &  $12.0 \pm 0.5$ & $5.95 \pm 0.25$ & $5.17$\\
CWW 89A & $5.6 \pm 0.5$ & $8.48 \pm 0.76$ & $5.29$ \\
KOI-607 & $9.1 \pm 0.5$ & $4.97 \pm 0.27$ &   $5.89$\\
\enddata
\vspace{-2.0cm}
\end{deluxetable}

Based on the results from Equation \ref{eq:tau}, it is not surprising that we find relatively large eccentricities for our objects. From the argument of the circularization timescale alone, not enough time has passed for each host star to circularize its companion's orbit and the asynchronization of the orbital period to the stellar rotation period agrees with this conclusion. If the orbit of the companion has been circularized or has near-zero eccentricity, then we expect the rotation period and orbital period to have synchronized since we normally expect the orbit to synchronize or pseudo-synchronize (in particular, at periastron) well before orbital circularization \citep{mazeh08}.

\begin{figure}[!ht]
\centering
\includegraphics[width=0.425\textwidth, trim= {2.0cm 0.0cm 3.3cm 1.0cm}]{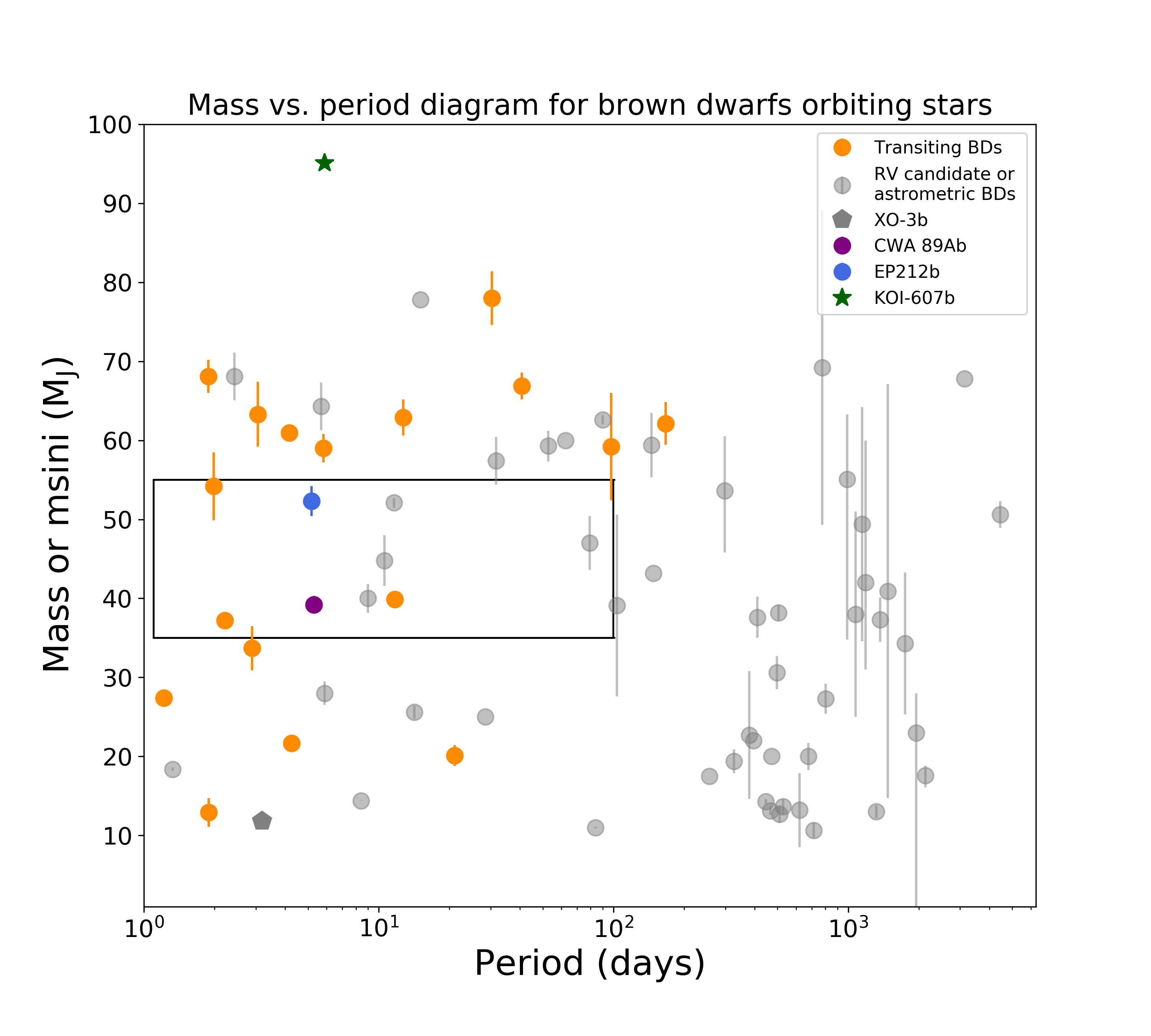}
\caption{Mass vs. Period plot adapted from \cite{ma12}. Note the gray colored points are minimum mass ($m\sin{i}$) values or masses derived from astrometry. The depleted region highlighted by \cite{ma12} is shown here in the black rectangle. Additional objects published after 2014 are shown here. A full list of the published transiting BDs as of March 2019 is shown in Table \ref{tab:bdlist}. See \cite{ma12} and \cite{csizmadia16} for a more comprehensive list that includes non-transiting BDs. \label{fig:gap}}
\end{figure}

\subsection{Defining Brown Dwarfs based on their formation mechanism}

The distinction between planets and stars is made clear, in part, by the formation mechanism of each group, so the same principle should be applied to BDs. Planets form out of the disks of developing stars via core accretion (and/or via disk instability) while stars form out of dense molecular clouds that collapse under gravity. There is no conclusive evidence that a $13\mj$ object (a nominal BD) cannot form via core accretion like giant planets while, likewise, there is no conclusive evidence that a more massive BD, say $60\mj$ cannot have formed in a different manner, i.e. like a star. If it were the case that these two BDs formed differently, then, we should define the BD that formed from core accretion instead as a giant planet since it formed as we believe giant planets form and abandon the arbitrary mass cutoff. The lower mass cutoff between BDs and planets needs to be motivated by the mass or mass range at which core accretion ends as a formation mechanism and cloud collapse begins as one.

In summary, the discovery of EP212b places a BD directly in the gap proposed by \cite{ma12} (see Figure \ref{fig:gap}), which serves as one more counter-example to the idea of two distinct BD populations. Moreover, a two-dimensional KS test cannot be used to determine if the population of BDs below $42.5\mj$ originates differently than the population above $42.5\mj$. From the circularization timescale discussion presented in \cite{adams06}, we find that EP212b, CWW 89Ab, and KOI-607b have not yet reached an age at which we expect their orbits to be circularized. This result is supported by the eccentric nature of each object's orbit as well as in the asynchronous characteristics between the host star's rotation period and the orbital period of the companion. We note again that we do not present AO imaging of EP212 to check for a third companion, but we believe that future studies should consider this since a third body may significantly effect the orbital motion of EP212b and alter our interpretation of the EP212 system.

EP212b is one of the only known transiting BDs at roughly $50\mj$ and is one of the first of more like it to be uncovered. Even more BDs may be awaiting discovery by the TESS mission where we may conduct transit searches on stars with companions of minimum mass measurements between $10\mj$ and $100\mj$. EP212b may be the first sign of a short-period BD population that leads us to question the presence of a gap in the mass distribution of BDs and is cause to more closely consider the nature of the brown dwarf desert.

\section{Acknowledgements}
We are grateful to the observers at the 1.5-meter telescope on Mt. Hopkins, Arizona for the TRES observations they took on our behalf. We also thank to \textit{Kepler} and \textit{K2} teams for the availability of the transit light curves of our targets. We also thank the referee for their thoughtful and thorough comments. Andrew Vanderburg's work was performed under contract with the California Institute of Technology (Caltech)/Jet Propulsion Laboratory (JPL) funded by NASA through the Sagan Fellowship Program executed by the NASA Exoplanet Science Institute. Funding for this work is provided by the National Science Foundation Graduate Research Fellowship Program Fellowship (GRFP). 

\software{SPC \citep{spc},
EXOFASTv2 \citep{eastman13}}

\bibliographystyle{aasjournal}
\bibliography{astro_citations}

\newpage

\begin{deluxetable}{cccccccccc}
\tabletypesize{\footnotesize}
\tablewidth{0pt}

 \tablecaption{List of published transiting brown dwarfs as of March 2019. \label{tab:bdlist}}

 \tablehead{
 \colhead{Name} & \colhead{Period (days)} & \colhead{$\rm M_{BD}/M_J$} & \colhead{$\rm R_{BD}/R_J$}& \colhead{e} & \colhead{$\rm M_\star/\mst$} &\colhead{$\rm R_\star/\rst$}& \colhead{$\rm T_{eff} (K)$}&\colhead{[Fe/H]} &\colhead{Reference}}
 \rotate
 \startdata 
 LP 261-75b & 1.882 & $68.09 \pm 2.10$ & $0.898 \pm 0.015$ & $<0.007$ & $0.300 \pm 0.015$ & $0.313 \pm 0.005$ & $3100 \pm 50$ & - & \cite{irwin18}\\
 NLTT 41135b & 2.889 & $33.70 \pm 2.80$ &  $1.13 \pm 0.27$ & $<0.02$ & $0.188 \pm 0.026$ & $0.210 \pm 0.016$ & $3230 \pm 130$ & $−0.25 \pm 0.25$ & \cite{irwin10}\\
 LHS 6343c$\rm ^a$ & 12.713 & $62.90 \pm 2.3$ & $0.833 \pm 0.021$ & $0.056 \pm 0.032$ & $0.37^{A}$, $0.30^{B}$ & $0.378^{A}$, -$^{B}$ & - & $+0.02 \pm 0.19$ & \cite{johnson11_bd}\\
 KELT-1b & 1.218 & $27.38 \pm 0.93$ & $1.116 \pm 0.038$ & $0.01 \pm 0.01$ & $1.335 \pm 0.063$ & $1.471 \pm 0.045$ & $6516 \pm 49$ & $+0.052 \pm 0.079$ & \cite{kelt1b}\\
 HATS-70b & 1.888 & $12.9 \pm 1.8$ & $1.384 \pm 0.079$ & $<0.18$ & $1.780 \pm 0.12$ & $1.881 \pm 0.066$ & $7930 \pm 820$ & $0.041 \pm 0.107$ & \cite{hats70b}\\
 XO-3b$\rm ^b$ & 3.192 & $11.79 \pm 0.59$ & $1.217 \pm 0.073$ & $0.260 \pm 0.017$ & $1.213 \pm 0.066$ & $1.377 \pm 0.083$ & $6429 \pm 100$ & $−0.177 \pm 0.080$ & \cite{xo3b}\\
 WASP-30b & 4.157 & $60.96 \pm 0.89$ & $0.889 \pm 0.021$ & 0 (adopted) & $1.166 \pm 0.026$ & $1.295 \pm 0.019$ & $6201 \pm 97$ & $-0.08 \pm 0.10$ & \cite{wasp30b}\\
 WASP-128b & 2.209 & $37.19 \pm 0.85$ & $0.937 \pm 0.022$ & $<0.007$ & $1.155 \pm 0.039$ & $1.152 \pm 0.021$ & $5950 \pm 50$ & $0.01 \pm 0.12$ & \cite{wasp128b}\\
 CoRoT-3b & 4.257 & $21.66 \pm 1.00$ & $1.010 \pm 0.070$ & 0 (adopted) & $1.370 \pm 0.090$ & $1.560 \pm 0.090$ & $6740 \pm 140$ & $-0.02 \pm 0.06$ & \cite{corot3b}\\
 CoRoT-15b & 3.060 & $63.30 \pm 4.10$ & $1.12 \pm 0.300$ & 0 (adopted) & $1.320 \pm 0.120$ & $1.460 \pm 0.310$ & $6350 \pm 200$ & $+0.10 \pm 0.20$ & \cite{corot15b}\\
 CoRoT-33b & 5.819 & $59.00 \pm 1.80$ & $1.10 \pm 0.53$ & $0.0700 \pm 0.0016$ & $0.86 \pm 0.04$ & $0.94 \pm 0.14$ & $5225 \pm 80$ & $+0.44 \pm 0.10$ & \cite{corot33b}\\
 Kepler-39b & 21.087 & $20.10 \pm 1.3$ & $1.24 \pm 0.10$ & $0.112 \pm 0.057$ & $1.29 \pm 0.07$ & $1.40 \pm 0.10$ & $6350 \pm 100$ & $+0.10 \pm 0.14$ & \cite{kepler39}\\
 KOI-189b$\rm ^c$ & 30.360 & $78.00 \pm 3.40$ & $0.998 \pm 0.023$ & $0.275 \pm 0.004$ & $0.764 \pm 0.051$ & $0.733 \pm 0.017$ & $4952 \pm 40$ & $-0.07 \pm 0.12$ & \cite{diaz14}\\
 KOI-205b & 11.720 & $39.9 \pm 1.0$ & $0.807 \pm 0.022$ & $<0.031$ & $0.925 \pm 0.033$ & $0.841 \pm 0.020$ & $5237 \pm 60$ & $+0.14 \pm 0.12$ & \cite{diaz13}\\
 KOI-415b & 166.788 & $62.14 \pm 2.69$ & $0.79 \pm 0.12$ & $0.689 \pm 0.0002$ & $0.94 \pm 0.06$ & $1.15 \pm 0.15$ & $5810 \pm 80$ & $-0.24 \pm 0.11$ & \cite{moutou13}\\
 KOI-607b$\rm ^d$ & 5.894 & $95.1 \pm 3.4$ & $1.089 \pm 0.089$ & $0.3950 \pm 0.0091$ & $0.993 \pm 0.052$ & $0.915 \pm 0.031$ & $5418 \pm 87$ & $0.376 \pm 0.088$ & this work\\
 EPIC 201702477b & 40.737 & $66.9 \pm 1.7$ & $0.757 \pm 0.065$ & $0.2281 \pm 0.0026$ & $0.870 \pm 0.031$ & $0.901 \pm 0.057$ & $5517 \pm 70$ & $-0.164 \pm 0.053$ & \cite{bayliss16}\\
 EPIC 212036875b & 5.170 & $52.3 \pm 1.9$ & $0.874 \pm 0.017$ & $0.1323 \pm 0.0042$ & $1.288 \pm 0.065$ & $1.498 \pm 0.026$ & $6238 \pm 60$ & $0.007 \pm 0.096$ & this work\\
 CWW 89Ab & 5.293 & $39.21 \pm 1.1$ & $0.941 \pm 0.019$ & $0.1891 \pm 0.0022$ & $1.101 \pm 0.045$ & $1.029 \pm 0.016$ & $5755 \pm 49$ & $0.203 \pm 0.091$ & this work\\
 AD 3116b $\rm ^e$ & 1.983 & $54.2 \pm 4.3$ & $1.02 \pm 0.28$ & $0.146 \pm 0.024$ & $0.276 \pm 0.020$ & $0.29 \pm 0.08$ & $3200 \pm 200$ & $0.16 \pm 0.1$ & \cite{ad3116} \\
 RIK 72b$\rm ^f$ & 97.76 & $59.2 \pm 6.8$ & $3.10 \pm 0.31$ & $0.146 \pm 0.0116$ & $0.439 \pm 0.044$ & $0.961 \pm 0.096$ & $3349 \pm 142$ & - & \cite{david19_bd}\\
 2M0535-05a$\rm ^g$ & 9.779 & $56.7 \pm 4.8$ & $6.50 \pm 0.33$ & $0.3225 \pm 0.0060$ & - & - & - & - & \cite{2M0535}\\
 2M0535-05b$\rm ^g$ & 9.779 & $35.6 \pm 2.8$ & $5.0 \pm 0.25$ & $0.3225 \pm 0.0060$ & - & - & - & - & \cite{2M0535}\\
 \enddata 
 \tablecomments{a - LHS 6343 is a triple system with two M-dwarfs and one brown dwarf, b - XO-3b is most likely a high mass gas giant exoplanet rather than a brown dwarf, c - KOI-189b is either a high mass brown dwarf or a very low mass star as its mass is near the substellar upper mass limit of $\rm 80M_J$, d - KOI-607b is a low mass star and not a brown dwarf, e - AD 3116b is a transiting brown dwarf associated with the Praesepe open cluster, f - RIK 72b is in the Upper Scorpius OB association}, g - 2M0535-05 is a brown dwarf binary.

\end{deluxetable}

\newpage

\begin{deluxetable}{ccccc}

\tabletypesize{\footnotesize}
\tablewidth{0pt}

 \tablecaption{Additional information on published transiting brown dwarfs. \label{tab:bdlist_extra}}

 \tablehead{
 \colhead{Name} & \colhead{$\alpha_{\rm J2000}$} & \colhead{$\delta_{\rm J2000}$} & \colhead{V (magnitude)}& \colhead{Reference}}

 \startdata 
 LP 261-75  & 09 51 04.58 & +35 58 09.47 & 15.33 & SIMBAD$\rm ^a$\\
 NLTT 41135 & 15 46 04.30 & +04 41 30.06 & 18.00 & \cite{irwin10}\\
 LHS 6343   & 19 10 14.28 & +46 57 24.11 & 13.38 & SIMBAD$\rm ^a$\\
 KELT-1     & 00 01 26.92 & +39 23 01.70 & 10.70 & \cite{kelt1b}\\
 HATS-70    & 07 16 25.08 & $-$31 14 39.86 & 12.57 & \cite{hats70b}\\
 XO-3$\rm ^b$ & 04 21 52.60 & +57 49 01.87 & 9.80 & \cite{xo3b}\\
 WASP-30    & 23 53 38.03 & $-$10 07 05.10 & 12.00 & \cite{wasp30b}\\
 WASP-128   & 11 31 26.10 & $-$41 41 22.30 & 12.50 & \cite{wasp128b}\\
 CoRoT-3    & 19 28 13.26 & +00 07 18.70 & 13.29 & \cite{corot3b}\\
 CoRoT-15   & 06 28 27.82 & +06 11 10.47 & 16.00 & \cite{corot15b}\\
 CoRoT-33   & 18 38 33.91 & +05 37 28.97 & 14.70 & \cite{corot33b}\\
 Kepler-39  & 19 47 50.46 & +46 02 03.49 & 14.47 & \cite{bouchy11}\\
 KOI-189$\rm ^c$ & 18 59 31.19	& +49 16 01.17 & 14.74 & \cite{diaz14}\\
 KOI-205    & 19 41 59.20 & +42 32 16.41 & 14.85 & \cite{diaz13}\\
 KOI-415    & 19 33 13.45	& +41 36 22.93 & 14.34 & \cite{moutou13}\\
 KOI-607$\rm ^d$ & 19 19 14.18 & +40 36 57.03 & 14.60 & this work\\
 EPIC 201702477 & 11 40 57.79 & +03 40 53.70 & 14.57 & \cite{bayliss16}\\
 EPIC 212036875 & 08 58 45.67 & +20 52 08.73 & 10.95 & this work\\
 CWW 89A    & 19 17 34.04 & $-$16 52 17.80 & 12.54 & this work\\
 AD 3116    & 08 42 39.43 & +19 24 51.90 & 18.73 & \cite{ad3116}\\
 RIK 72     & 16 03 39.22 & $-$18 51 29.72 & 14.35G$\rm ^e$ & \cite{david19_bd} \\
 2M0535-05 & 05 35 21.85 & $-$05 46 08.56 & 18.94G$\rm ^e$ & \cite{2M0535}\\
 \enddata
 \tablecomments{a - SIMBAD database: \url{http://simbad.u-strasbg.fr/simbad/}, b - XO-3b is most likely a high mass gas giant exoplanet rather than a brown dwarf, c - KOI-189b is either a high mass brown dwarf or a very low mass star as its mass is near the substellar upper mass limit of $\rm 80M_J$, d - KOI-607b is a low mass star and not a brown dwarf, e - G-band magnitude from the \textit{Gaia} mission.}

\end{deluxetable}

\newpage

\begin{deluxetable}{lcc}
\tablecaption{Median values and 68\% confidence interval for CWW 89Ab, created using EXOFASTv2 commit number 65aa674. \label{tab:EP219388192_MIST}}
\tablehead{\colhead{~~~Parameter} & \colhead{Units} & \multicolumn{1}{c}{Values}}
\startdata
\smallskip\\\multicolumn{2}{l}{Stellar Parameters:}&\smallskip\\
~~~~$M_\star$\dotfill &Mass (\msun)\dotfill &$1.101^{+0.039}_{-0.045}$\\
~~~~$R_\star$\dotfill &Radius (\rsun)\dotfill &$1.029\pm0.016$\\
~~~~$L_\star$\dotfill &Luminosity (\lsun)\dotfill &$1.048\pm0.043$\\
~~~~$\rho_\star$\dotfill &Density (cgs)\dotfill &$1.423\pm0.076$\\
~~~~$\log{g}$\dotfill &Surface gravity (cgs)\dotfill &$4.455^{+0.018}_{-0.020}$\\
~~~~$T_{\rm eff}$\dotfill &Effective Temperature (K)\dotfill &$5755\pm49$\\
~~~~$[{\rm Fe/H}]$\dotfill &Metallicity (dex)\dotfill &$0.203^{+0.086}_{-0.091}$\\
~~~~$Age$\dotfill &Age (Gyr)\dotfill &$1.26^{+1.7}_{-0.93}$\\
~~~~$A_V$\dotfill &V-band extinction (mag)\dotfill &$0.048^{+0.068}_{-0.035}$\\
~~~~$\varpi$\dotfill &Parallax (mas)\dotfill &$3.304\pm0.046$\\
~~~~$d$\dotfill &Distance (pc)\dotfill &$302.7\pm4.2$\\
\smallskip\\\multicolumn{2}{l}{Brown Dwarf Parameters:}&b\smallskip\\
~~~~$P$\dotfill &Period (days)\dotfill &$5.292600^{+0.000018}_{-0.000017}$\\
~~~~$R_B$\dotfill &Radius (\rj)\dotfill &$0.941\pm0.019$\\
~~~~$T_C$\dotfill &Time of conjunction (\bjdtdb)\dotfill &$2457341.037107^{+0.000087}_{-0.000090}$\\
~~~~$T_0$\dotfill &Optimal conjunction Time (\bjdtdb)\dotfill &$2457346.329706^{+0.000087}_{-0.000089}$\\
~~~~$a$\dotfill &Semi-major axis (AU)\dotfill &$0.06206^{+0.00071}_{-0.00084}$\\
~~~~$i$\dotfill &Inclination (Degrees)\dotfill &$88.53^{+0.27}_{-0.24}$\\
~~~~$e$\dotfill &Eccentricity \dotfill &$0.1891^{+0.0021}_{-0.0022}$\\
~~~~$\omega_\star$\dotfill &Argument of Periastron (Degrees)\dotfill &$-13.6^{+1.3}_{-1.2}$\\
~~~~$M_B$\dotfill &Mass (\mj)\dotfill &$39.21^{+0.91}_{-1.1}$\\
~~~~$K$\dotfill &RV semi-amplitude (m/s)\dotfill &$4269^{+18}_{-17}$\\
~~~~$R_B/R_\star$\dotfill &Radius of planet in stellar radii \dotfill &$0.09394^{+0.00057}_{-0.00058}$\\
~~~~$a/R_\star$\dotfill &Semi-major axis in stellar radii \dotfill &$12.96\pm0.23$\\
~~~~$\delta$\dotfill &Transit depth (fraction)\dotfill &$0.00883\pm0.00011$\\
~~~~$Depth$\dotfill &Flux decrement at mid transit \dotfill &$0.00883\pm0.00011$\\
~~~~$\tau$\dotfill &Ingress/egress transit duration (days)\dotfill &$0.01336^{+0.00061}_{-0.00056}$\\
~~~~$T_{14}$\dotfill &Total transit duration (days)\dotfill &$0.13928^{+0.00053}_{-0.00051}$\\
~~~~$b$\dotfill &Transit Impact parameter \dotfill &$0.335^{+0.049}_{-0.058}$\\
~~~~$\rho_B$\dotfill &Density (cgs)\dotfill &$58.3^{+3.9}_{-3.6}$\\
~~~~$logg_B$\dotfill &Surface gravity \dotfill &$5.040\pm0.020$\\
~~~~$T_P$\dotfill &Time of Periastron (\bjdtdb)\dotfill &$2457339.832^{+0.019}_{-0.018}$\\
~~~~$ecos{\omega_*}$\dotfill & \dotfill &$0.1837\pm0.0021$\\
~~~~$esin{\omega_*}$\dotfill & \dotfill &$-0.0445^{+0.0043}_{-0.0039}$\\
~~~~$M_B\sin i$\dotfill &Minimum mass (\mj)\dotfill &$39.19^{+0.91}_{-1.1}$\\
~~~~$M_B/M_\star$\dotfill &Mass ratio \dotfill &$0.03399^{+0.00051}_{-0.00042}$\\
\smallskip\\\multicolumn{2}{l}{Wavelength Parameters:}&Kepler\smallskip\\
~~~~$u_{1}$\dotfill &linear limb-darkening coeff \dotfill &$0.414\pm0.024$\\
~~~~$u_{2}$\dotfill &quadratic limb-darkening coeff \dotfill &$0.191^{+0.046}_{-0.045}$\\
\smallskip\\\multicolumn{2}{l}{Telescope Parameters:}&TRES\smallskip\\
~~~~$\gamma_{\rm rel}$\dotfill &Relative RV Offset (m/s)\dotfill &$3489\pm13$\\
~~~~$\sigma_J$\dotfill &RV Jitter (m/s)\dotfill &$16^{+24}_{-16}$\\
~~~~$\sigma_J^2$\dotfill &RV Jitter Variance \dotfill &$260^{+1400}_{-700}$\\
\enddata
\end{deluxetable}

\newpage

\begin{deluxetable}{lcc}
\tablecaption{Median values and 68\% confidence interval for EPIC 212036875b, created using EXOFASTv2 commit number 65aa674. \label{tab:EP212036875_MIST}}
\tablehead{\colhead{~~~Parameter} & \colhead{Units} & \multicolumn{1}{c}{Values}}
\startdata
\smallskip\\\multicolumn{2}{l}{Stellar Parameters:}&\smallskip\\
~~~~$M_\star$\dotfill &Mass (\msun)\dotfill &$1.288^{+0.065}_{-0.064}$\\
~~~~$R_\star$\dotfill &Radius (\rsun)\dotfill &$1.498^{+0.025}_{-0.026}$\\
~~~~$L_\star$\dotfill &Luminosity (\lsun)\dotfill &$3.06^{+0.15}_{-0.14}$\\
~~~~$\rho_\star$\dotfill &Density (cgs)\dotfill &$0.539^{+0.030}_{-0.027}$\\
~~~~$\log{g}$\dotfill &Surface gravity (cgs)\dotfill &$4.196\pm0.020$\\
~~~~$T_{\rm eff}$\dotfill &Effective Temperature (K)\dotfill &$6238^{+59}_{-60}$\\
~~~~$[{\rm Fe/H}]$\dotfill &Metallicity (dex)\dotfill &$0.007\pm0.096$\\
~~~~$Age$\dotfill &Age (Gyr)\dotfill &$2.70^{+0.98}_{-0.84}$\\
~~~~$A_V$\dotfill &V-band extinction (mag)\dotfill &$0.027^{+0.021}_{-0.019}$\\
~~~~$\varpi$\dotfill &Parallax (mas)\dotfill &$3.257\pm0.045$\\
~~~~$d$\dotfill &Distance (pc)\dotfill &$307.1^{+4.3}_{-4.2}$\\
\smallskip\\\multicolumn{2}{l}{Brown Dwarf Parameters:}&b\smallskip\\
~~~~$P$\dotfill &Period (days)\dotfill &$5.169885^{+0.000027}_{-0.000026}$\\
~~~~$R_B$\dotfill &Radius (\rj)\dotfill &$0.874\pm0.017$\\
~~~~$T_C$\dotfill &Time of conjunction (\bjdtdb)\dotfill &$2458129.69863\pm0.00014$\\
~~~~$T_0$\dotfill &Optimal conjunction Time (\bjdtdb)\dotfill &$2458134.86852^{+0.00013}_{-0.00014}$\\
~~~~$a$\dotfill &Semi-major axis (AU)\dotfill &$0.0645\pm0.0011$\\
~~~~$i$\dotfill &Inclination (Degrees)\dotfill &$83.93\pm0.16$\\
~~~~$e$\dotfill &Eccentricity \dotfill &$0.1323^{+0.0042}_{-0.0041}$\\
~~~~$\omega_\star$\dotfill &Argument of Periastron (Degrees)\dotfill &$160.5^{+5.2}_{-5.0}$\\
~~~~$M_B$\dotfill &Mass (\mj)\dotfill &$52.3\pm1.9$\\
~~~~$K$\dotfill &RV semi-amplitude (m/s)\dotfill &$5078\pm57$\\
~~~~$R_B/R_\star$\dotfill &Radius of planet in stellar radii \dotfill &$0.05998^{+0.00069}_{-0.00070}$\\
~~~~$a/R_\star$\dotfill &Semi-major axis in stellar radii \dotfill &$9.25^{+0.17}_{-0.16}$\\
~~~~$\delta$\dotfill &Transit depth (fraction)\dotfill &$0.003597^{+0.000083}_{-0.000084}$\\
~~~~$Depth$\dotfill &Flux decrement at mid transit \dotfill &$0.003597^{+0.000083}_{-0.000084}$\\
~~~~$\tau$\dotfill &Ingress/egress transit duration (days)\dotfill &$0.0285\pm0.0015$\\
~~~~$T_{14}$\dotfill &Total transit duration (days)\dotfill &$0.08932^{+0.00061}_{-0.00059}$\\
~~~~$b$\dotfill &Transit Impact parameter \dotfill &$0.9207^{+0.0040}_{-0.0047}$\\
~~~~$\rho_B$\dotfill &Density (cgs)\dotfill &$97.1^{+5.2}_{-5.0}$\\
~~~~$\log{g_B}$\dotfill &Surface gravity \dotfill &$5.229\pm0.017$\\
~~~~$T_P$\dotfill &Time of Periastron (\bjdtdb)\dotfill &$2458130.513^{+0.064}_{-0.062}$\\
~~~~$ecos{\omega_*}$\dotfill & \dotfill &$-0.1243^{+0.0064}_{-0.0063}$\\
~~~~$esin{\omega_*}$\dotfill & \dotfill &$0.044^{+0.010}_{-0.011}$\\
~~~~$M_B\sin i$\dotfill &Minimum mass (\mj)\dotfill &$52.0\pm1.9$\\
~~~~$M_B/M_\star$\dotfill &Mass ratio \dotfill &$0.03876^{+0.00082}_{-0.00079}$\\
\smallskip\\\multicolumn{2}{l}{Wavelength Parameters:}&Kepler\smallskip\\
~~~~$u_{1}$\dotfill &linear limb-darkening coeff \dotfill &$0.312^{+0.047}_{-0.049}$\\
~~~~$u_{2}$\dotfill &quadratic limb-darkening coeff \dotfill &$0.292\pm0.048$\\
\smallskip\\\multicolumn{2}{l}{Telescope Parameters:}&TRES\smallskip\\
~~~~$\gamma_{\rm rel}$\dotfill &Relative RV Offset (m/s)\dotfill &$5691^{+49}_{-51}$\\
~~~~$\sigma_J$\dotfill &RV Jitter (m/s)\dotfill &$0.00^{+84}_{-0.00}$\\
~~~~$\sigma_J^2$\dotfill &RV Jitter Variance \dotfill &$-1900^{+8900}_{-2800}$\\
\enddata
\end{deluxetable}

\newpage

\begin{deluxetable}{lcc}
\tablecaption{Median values and 68\% confidence interval for KOI-607b, created using EXOFASTv2 commit number 65aa674. \label{tab:K00607_MIST}}
\tablehead{\colhead{~~~Parameter} & \colhead{Units} & \multicolumn{1}{c}{Values}}
\startdata
\smallskip\\\multicolumn{2}{l}{Stellar Parameters:}&\smallskip\\
~~~~$M_\star$\dotfill &Mass (\msun)\dotfill &$0.993^{+0.050}_{-0.052}$\\
~~~~$R_\star$\dotfill &Radius (\rsun)\dotfill &$0.915^{+0.031}_{-0.028}$\\
~~~~$L_\star$\dotfill &Luminosity (\lsun)\dotfill &$0.651^{+0.067}_{-0.061}$\\
~~~~$\rho_\star$\dotfill &Density (cgs)\dotfill &$1.82\pm0.14$\\
~~~~$\log{g}$\dotfill &Surface gravity (cgs)\dotfill &$4.511^{+0.021}_{-0.023}$\\
~~~~$T_{\rm eff}$\dotfill &Effective Temperature (K)\dotfill &$5418^{+87}_{-85}$\\
~~~~$[{\rm Fe/H}]$\dotfill &Metallicity (dex)\dotfill &$0.376^{+0.075}_{-0.088}$\\
~~~~$Age$\dotfill &Age (Gyr)\dotfill &$2.6^{+3.4}_{-1.9}$\\
~~~~$A_V$\dotfill &V-band extinction (mag)\dotfill &$0.101^{+0.12}_{-0.074}$\\
~~~~$\varpi$\dotfill &Parallax (mas)\dotfill &$1.489\pm0.056$\\
~~~~$d$\dotfill &Distance (pc)\dotfill &$671^{+26}_{-24}$\\
\smallskip\\\multicolumn{2}{l}{Brown Dwarf Parameters:}&b\smallskip\\
~~~~$P$\dotfill &Period (days)\dotfill &$5.89399148\pm0.00000060$\\
~~~~$R_B$\dotfill &Radius (\rj)\dotfill &$1.089^{+0.089}_{-0.061}$\\
~~~~$T_C$\dotfill &Time of conjunction (\bjdtdb)\dotfill &$2455006.485529^{+0.000086}_{-0.000090}$\\
~~~~$T_0$\dotfill &Optimal conjunction Time (\bjdtdb)\dotfill &$2455648.930598^{+0.000057}_{-0.000060}$\\
~~~~$a$\dotfill &Semi-major axis (AU)\dotfill &$0.0656^{+0.0010}_{-0.0011}$\\
~~~~$i$\dotfill &Inclination (Degrees)\dotfill &$84.61^{+0.24}_{-0.26}$\\
~~~~$e$\dotfill &Eccentricity \dotfill &$0.3950^{+0.0091}_{-0.0090}$\\
~~~~$\omega_\star$\dotfill &Argument of Periastron (Degrees)\dotfill &$116.0\pm1.3$\\
~~~~$M_B$\dotfill &Mass (\mj)\dotfill &$95.1^{+3.3}_{-3.4}$\\
~~~~$K$\dotfill &RV semi-amplitude (m/s)\dotfill &$10990^{+140}_{-130}$\\
~~~~$R_B/R_\star$\dotfill &Radius of planet in stellar radii \dotfill &$0.1221^{+0.0065}_{-0.0037}$\\
~~~~$a/R_\star$\dotfill &Semi-major axis in stellar radii \dotfill &$15.39^{+0.38}_{-0.40}$\\
~~~~$\delta$\dotfill &Transit depth (fraction)\dotfill &$0.01418^{+0.00045}_{-0.00042}$\\
~~~~$\tau$\dotfill &Ingress/egress transit duration (days)\dotfill &$0.02780^{+0.00030}_{-0.00034}$\\
~~~~$T_{14}$\dotfill &Total transit duration (days)\dotfill &$0.05563\pm0.00058$\\
~~~~$b$\dotfill &Transit Impact parameter \dotfill &$0.900^{+0.017}_{-0.013}$\\
~~~~$\rho_B$\dotfill &Density (cgs)\dotfill &$91^{+16}_{-19}$\\
~~~~$\log{g_B}$\dotfill &Surface gravity \dotfill &$5.299^{+0.046}_{-0.065}$\\
~~~~$ecos{\omega_\star}$\dotfill & \dotfill &$-0.1732^{+0.0071}_{-0.0073}$\\
~~~~$esin{\omega_\star}$\dotfill & \dotfill &$0.355\pm0.010$\\
~~~~$M_B\sin i$\dotfill &Minimum mass (\mj)\dotfill &$94.7^{+3.2}_{-3.4}$\\
~~~~$M_B/M_\star$\dotfill &Mass ratio \dotfill &$0.0915^{+0.0020}_{-0.0019}$\\
\smallskip\\\multicolumn{2}{l}{Wavelength Parameters:}&Kepler\smallskip\\
~~~~$u_{1}$\dotfill &linear limb-darkening coeff \dotfill &$0.484^{+0.052}_{-0.054}$\\
~~~~$u_{2}$\dotfill &quadratic limb-darkening coeff \dotfill &$0.189^{+0.050}_{-0.052}$\\
\smallskip\\\multicolumn{2}{l}{Telescope Parameters:}&TRES\smallskip\\
~~~~$\gamma_{\rm rel}$\dotfill &Relative RV Offset (m/s)\dotfill &$2224^{+61}_{-68}$\\
~~~~$\sigma_J$\dotfill &RV Jitter (m/s)\dotfill &$132^{+110}_{-90}$\\
~~~~$\sigma_J^2$\dotfill &RV Jitter Variance \dotfill &$18000^{+41000}_{-16000}$\\
\enddata
\end{deluxetable}

\end{document}